\documentclass[a4paper,12pt]{article}  
\usepackage[dvips]{epsfig}
\usepackage{amssymb}
\usepackage{cite}
\newcommand{\vev}[1]{\left\langle #1\right\rangle}

\newcommand{\goes}{\rightarrow} 
 
\newcommand{\GeV}{\; \mathrm{GeV}} 
\newcommand{\TeV}{\; \mathrm{TeV}} 
\newcommand{\lapproxeq}{\lower .7ex\hbox{$\;\stackrel{\textstyle  
<}{\sim}\;$}} 
\newcommand{\gapproxeq}{\lower .7ex\hbox{$\;\stackrel{\textstyle  
>}{\sim}\;$}} 
\newcommand{\stackdown}[2]{\lower 1.4ex\hbox{$\;\stackrel{\textstyle{#1}}  
{\scriptstyle{#2}}\;$}}
\newcommand{\beq}{\begin{equation}} 
\newcommand{\eeq}{\end{equation}} 
\newcommand{\bea}{\begin{eqnarray}} 
\newcommand{\eea}{\end{eqnarray}}
\newcommand{\lsp}{\tilde{\chi}}
\newcommand{\mlsp}{m_{\lsp}}
\newcommand{\xsec}{\vev{ \sigma v_{rel} }}

\newcommand{\relic}{\Omega_{\lsp}\,h_0^2} 
\newcommand{\reaction}{\lsp\,\lsp \goes X\,Y} 
\newcommand{\etal}{\textit{et. al.}}

\makeatletter 
\def\slash{\@ifnextchar[{\fmsl@sh}{\fmsl@sh[0mu]}} 
\def\fmsl@sh[#1]#2{% 
  \mathchoice 
    {\@fmsl@sh\displaystyle{#1}{#2}}% 
    {\@fmsl@sh\textstyle{#1}{#2}}% 
    {\@fmsl@sh\scriptstyle{#1}{#2}}% 
    {\@fmsl@sh\scriptscriptstyle{#1}{#2}}} 
\def\@fmsl@sh#1#2#3{\m@th\ooalign{$\hfil#1\mkern#2/\hfil$\crcr$#1#3$}} 
\makeatother 
\begin{document} 
\begin{titlepage} 
 
%%%%%%%%%%% 
\begin{flushright} 
\parbox{6.6cm}{ ACT-5/99, CTP-TAMU-25/99 \\ UA/NPPS-03-99, 
                         hep-ph/9906394 } 
\end{flushright} 
%%%%%%%%%% 
\begin{centering} 
\vspace*{1.5cm} 
 
{\large{\textbf {Neutralino Relic Density with a Cosmological Constant 
confronts Electroweak Precision Measurements}}}\\ 
\vspace{1.4cm} 
 
{\bf A.~B.\ Lahanas} $^{1}$, \, 
{\bf D.~V.~Nanopoulos} $^{2}$  \, and \, {\bf V.~C.~Spanos} $^{1}$  \\ 
\vspace{.8cm} 
$^{1}$ {\it University of Athens, Physics Department,  
Nuclear and Particle Physics Section,\\  
GR--15771  Athens, Greece}\\ 
 
\vspace{.5cm} 
$^{2}$ {\it Department of Physics,  
         Texas A \& M University, College Station,  
         TX~77843-4242, USA, 
         Astroparticle Physics Group, Houston 
         Advanced Research Center (HARC), Mitchell Campus, 
         Woodlands, TX 77381, USA, and \\ 
         Academy of Athens,  
         Chair of Theoretical Physics,  
         Division of Natural Sciences, 28~Panepistimiou Avenue,  
         Athens 10679, Greece}  \\ 
\end{centering} 
\vspace{3.cm} 
%%%%%%%%%%%%%%%%%%%%%%%%% 
\begin{abstract} 
We discuss the relic density of the lightest of the  
supersymmetric particles ({\small LSP})   
in view of new cosmological data, which favour the  
concept of an accelerating 
Universe with a non-vanishing cosmological constant. The new bound on the 
Cold Dark Matter  
density, $\Omega_{\mathrm{CDM}} h_0^2 \lesssim 0.22$, 
puts stringent constraints on supersymmetry preferring low 
supersymmetry breaking scales, 
in sharp contrast to electroweak precision measurements 
favouring large supersymmetry breaking scales.  
Supersymmetric predictions are in agreement with cosmological 
data and electroweak precision data 
in the window of the parameter space:  
$m_0<200\GeV$, $300\GeV<M_{1/2}<400\GeV$, putting  
bounds on sparticle masses, which may be evaded if
$m_{_{\mathrm{LSP}}}<m_{{\tilde{\tau}}_R} \lesssim 1.2 \; m_{_{\mathrm{LSP}}}$.
\end{abstract} 
\end{titlepage} 
\newpage 
\baselineskip=20pt 

%%%%%%%%%%%%%%%%%%%%%%%%%% Letter body %%%%%%%%%%%%%%%%%%%%%%%%%    
Recent observations of type Ia supernovae ({\small SNI}a)
put new constraints on
the cosmological parameters. The data favour
an almost {\em flat} and {\em accelerating} Universe,
where the acceleration mainly is driven by a {\em non-vanishing
cosmological constant}.

There is a growing consensus that the anisotropy of
the Cosmic Background Radiation ({\small CBR}) offers the best
way to determine the curvature of the Universe and hence the total
matter-energy density $\Omega_0$ \cite{Turner}.
The data are consistent
with a flat Universe, since ${\Omega}_0=1.0 \pm 0.2$
\cite{Turner,Lineweaver}, and
the radiation content of the matter-energy density,
that is contribution coming from 
{\small CBR} and/or
ultra relativistic neutrinos,
is very small. 
Therefore the present matter-energy density
can be decomposed basically into two components: the matter density
$\Omega_{\mathrm{M}}$ and the vacuum energy $\Omega_{\Lambda}$:
\beq
\Omega_0=\Omega_{\mathrm{M}}+\Omega_{\Lambda}\,. \label{omega0}
\eeq

There is supporting evidence, coming 
from many independent astrophysical observations, 
that the matter density weighs $\Omega_{\mathrm{M}}=0.4 \pm 0.1$.
Recently two groups, the Supernova Cosmology Project \cite{Perlmutter}
and the High-$z$ Supernova Search Team \cite{Riess},
using different methods of analysis, each found evidence
for accelerated expansion, driven by a vacuum energy contribution:
\beq
\Omega_{\Lambda}={4\over{3}}\Omega_{\mathrm{M}}+{1\over{3}}\pm {1\over{6}}\,.
\eeq
So, for $\Omega_{\mathrm{M}}=0.4 \pm 0.1$ this relation implies that
the vacuum energy is non-vanishing,
$\Omega_{\Lambda}=0.85\pm 0.2$, 
value which is compatible with
a flat Universe, as the anisotropy of 
{\small CBR} measurements indicate.
Taking into account the fact that 
the baryonic contribution to the matter density
is small, ${\Omega}_{\mathrm{B}}=0.05 \pm 0.005 $,
the values for matter energy density $\Omega_{\mathrm{M}}$ 
result to a Cold Dark Matter ({\small CDM}) density 
${\Omega}_{\mathrm{CDM}} \simeq 0.35 \pm 0.1$, which combined with 
more recent measurements \cite{Turner,Freedman}
of the scaled Hubble parameter
$h_0=0.65 \pm 0.05$,  result to small
{\small CDM} relic densities:  
\beq 
\Omega_{\mathrm{CDM}} \, {h_0}^2 \simeq 0.15 \pm 0.07\,.
\label{bound}
\eeq 

Such stringent bounds for the {\small CDM} relic density
do affect supersymmetric
predictions and can severely lower the limits of the
effective supersymmetry
breaking scale and hence
the masses of the supersymmetric particles,
as first emphasized in Ref.~\cite{Lopez0}.
The {\small CDM} relic density with non-vanishing
cosmological constant in the framework of the
Minimal Supersymmetric Standard Model ({\small MSSM}) has
been the subject of Ref.~\cite{Wells}.
Using the recent cosmological data, gauge fermions are
predicted to be within {\small LHC} reach. 

It is perhaps worth pointing out 
that while electroweak ({\small EW}) precision data are in perfect
agreement with Standard Model ({\small SM}) predictions, and
hence in agreement too with
supersymmetric models characterized by a large
supersymmetry breaking scale
${M}_{\mathrm{SUSY}}$ \cite{Lahanas}, the data on
${\Omega}_{\mathrm{CDM}} \, {h_0}^2$
push ${M}_{\mathrm{SUSY}}$
 to the opposite direction favouring  small values of
${M}_{\mathrm{SUSY}}$.
Therefore {\small EW} precision data may not reconcile with
the assumption that the lightest supersymmetric
 particle ({\small LSP} or $\lsp$), is a candidate for
{\small CDM}.

In $R$-parity conserving supersymmetric 
theories the {\small LSP} 
is stable, and for most of supersymmetric models is the lightest
neutralino, which is a good candidate for the {\small CDM} particle \cite{first}.
Many authors 
\cite{old1,Sred,griest2,old2,old3,Mizuta,Lopez2,Lopez3,arno,Lopez,Drees,%
recent,reports,Drees2,Falk1,Falk2}
have calculated the relic neutralino density.  
In the early works, only the most important neutralino annihilation
channels were considered, but later works \cite{Drees,recent} included all
annihilation channels.
Also more refined calculations of thermal averages of cross sections 
were employed, which took into account threshold effects and integration over
Breit--Wigner poles \cite{Griest,Gelmini}.

Our study in this letter is based on the constrained
supergravity ({\small SUGRA}) scenario, assuming
universal boundary conditions for
the soft breaking parameters 
at the unification 
scale $M_{\mathrm{GUT}}$\footnote{We allow however
for small deviations from the gauge
coupling unification scenario. In this case the value of the strong
coupling constant at the unification scale, defined as the point where
the couplings $\alpha_1$ and $\alpha_2$ meet, is different
from $\alpha_{1,2}$.}.
It is also assumed that the {\small EW} symmetry is 
radiatively broken \cite{report}.
Therefore the arbitrary  
parameters are: $m_0$, $M_{1/2}$,  $A_0$ and $\tan \beta$.  
The absolute value of $\mu$ is  
determined from the minimization conditions of the one-loop corrected  
effective potential. These also determine  
the Higgs mixing parameter $m_3^2$. The sign of $\mu$ is undetermined  
in this procedure and in our analysis both signs of $\mu$ are considered.  
Therefore in this scheme the $\mu$ value as well as $m_3^2 \equiv B \mu$  
are not inputs.  

The Boltzmann transport equation for the neutralino 
number density is:
\beq
 {{dn}\over{dt}} =-3H n -\xsec (n^2(T)-n^2_0(T)) \,,
\eeq
where $n$ is the number of $\lsp$'s per unit volume, $n_0$ is 
their density in thermal equilibrium and $H$ is the
expansion rate of the Universe.
Using the fact that the total entropy 
$S=h\,T^3\,R^3$ is conserved and defining the 
quantity $q(x) \equiv \frac{n(T)}{T^3 h(T)}$ with $x\equiv T/\mlsp$, 
this differential equation can be cast into a form suitable for
numerical manipulations \cite{Sred}:
\beq 
\frac{dq}{dx} = \lambda (x) \; (q^2 - q_0^2) \,, \label{Bol}
\eeq 
where
\beq 
\lambda (x) \equiv 
  \left( {{\frac{4 \pi^3}{45}} G_N} \right)^{-1/2}  
 {\frac{m_{\lsp}}{\sqrt{g(T)}}}  
\left( h(T)+ {\frac{m_{\lsp}}{3}}{{h}^{\prime}}(T) \right) \xsec \, .  \label{lambda}
\eeq 
The functions $g(T)$ and $h(T)$ appearing in the equation above,  
are the effective energy  and entropy degrees of freedom respectively,
and they determine the 
Universe energy and entropy density through the relations:
\beq
\rho(T) \, = \, {\frac{\pi^2}{30}} \, T^4 \; g(T)\;,\;\;
s(T) = {\frac{2 \pi^2}{45}} \; T^3 \, h(T)\; .  
\eeq
Solving Eq. (\ref{Bol}) we obtain the $q(x_0)$, where $x_0 \equiv T_0/\mlsp$
corresponds to today's Universe temperature 
$T_0 \approx 2.7 \;^0\mathrm{K} $.
Using that $q \equiv \frac{n(T)}{T^3 h(T)}$,
$\Omega_{\lsp} \;=\; \frac{\rho_{\lsp}}{\rho_{\mathrm{crit}}}$ and
$\rho_{\mathrm{crit}}={3H_0^2\over{8\pi G_N}}$, we determine the
present value of neutralino relic density $\relic$ from 
the equation: 
\beq
\relic= {8\pi \over{3}}\; T_0^3\;G_N\; \mlsp \; h(x_0)\;q(x_0)\;
     (100\,\mathrm{km}\,\mathrm{sec}^{-1}\,\mathrm{Mpc}^{-1})^{-2}
       \, .
\eeq

Before solving the Boltzmann equation we need to calculate 
the effective degrees of freedom functions $g(T)$ and $h(T)$, 
as well as the thermally averaged cross section $\xsec$,
which enter into Eq. (\ref{Bol}).
Regarding the calculation of the functions $g(T)$ and $h(T)$,
the content of the particles in equilibrium is different depending on the
temperature $T$.
In our analyses we use the expressions for $g(T)$, $h(T)$ as
given in Ref.~\cite{Lopez}. 
Next we turn to the calculation of $\xsec$. 
At this point we follow
Ref.~\cite{Drees} and express the non-relativistic cross sections for the
various annihilation processes 
$\reaction$, in terms of helicity
amplitudes. We follow the standard treatment 
and ignore contributions of all channels, which are forbidden at zero relative 
velocity $v_{rel}$ of the two annihilating $\lsp$'s.
This approximation 
is not expected to invalidate significantly 
our results.  
So we  consider only the contributions of channels with non-supersymmetric
particles in the final state:
$$
q{\bar q},\; l{\bar l},\;
W^{+}W^{-},\; ZZ,\; ZH,\; Zh,\; ZA,\; W^{\pm}H^{\mp},\;  
HH,\; hh,\; Hh,\;AA,\; HA, \; hA,\; H^{+}H^{-}\,. 
$$
$q$ and $l$ denote quarks and leptons respectively, while
$H$, $h$, $A$ and $H^{\pm}$  denote the heavy, light, pseudoscalar
and  charged Higgses respectively.
In our analysis we have not studied neutralino--stau coannihilation effects 
\cite{Falk1,Griest,Mizuta}, which if included can lower the values of the 
neutralino relic density. Although important, 
coannihilation processes are of relevance for values of the parameters for
which $m_{\lsp}<m_{{\tilde{\tau}}_R} \lesssim 1.2 \; m_{\lsp} $, that is 
near the edge where $\lsp$ and
$\tilde{\tau}_{R}$ are almost degenerate in mass.

It is well known that the non-relativistic expansion in the relative velocity 
$v_{rel}$ breaks down near thresholds or poles of the cross sections, 
and in these cases, results based on this expansion are unreliable.
We locate the points of the parameter 
space of the {\small MSSM}, which result to pole and/or thresholds of the
cross section, using ``near pole" and ``near threshold" criteria.  
The comparison of our results with those of other 
studies \cite{Lopez3,recent}, which treat the problem of poles and thresholds 
in a more accurate manner, shows that they 
are in striking agreement. This occurs, at least, in regions of the parameter 
space of {\small MSSM} where this comparison is feasible.

Knowing  
$\xsec$, from the procedure outlined previously, and by calculating  
the functions $g(T)$, $h(T)$, $h^\prime (T)$  we can have the prefactor  
$\lambda (x)$ appearing in Eq. (\ref{Bol}).  
At high temperatures, or same  
large values of $x=T/\mlsp$ above the freeze-out temperature,  
the function $q(x)$ approaches its  
equilibrium value $q_0 (x)$ (see Eq. \ref{Bol}). 
A convenient and accurate method for  
solving the Boltzmann equation is the {\small WKB} 
approximation employed in Ref.~\cite{Lopez}.
As far as the scanning of the parameter space of the {\small MSSM}
is concerned, we exclude points that are theoretically 
forbidden, such as those leading to breaking of lepton and/or color number, 
or points for which Landau poles are developed and so on. We also exclude 
points for which the {\small LSP} is not a neutralino, as well as 
points of the parameter space for 
which violation of the experimental bounds on sparticle masses is 
encountered. We use the bounds which are listed in Ref.~\cite{data}.
From these bounds we have found that the chargino mass bound turns out to be
the most stringent one.
Details on our calculation will be published
elsewhere \cite{LNS}.

Before embarking to discuss our physics results we should stress that in our 
scheme we have not made any approximation concerning the masses or couplings 
of sectors which are rather involved such as neutralini for instance, which 
are crucial for our analysis. Therefore we do not only consider regions of the 
parameter space in which the {\small LSP} is purely $ \tilde B$ (bino) or
purely a Higgsino, but also regions where in general the {\small LSP} 
happens to be an admixture of the four available degrees of 
freedom\footnote{ The case of a Higgsino-like 
LSP has been pursued in Refs.~\cite{Drees2,Falk2},
where the dominant radiative corrections to neutralino 
masses are considered. Analogous corrections to couplings of Higgsino-like 
neutralinos to $Z$ and Higgs bosons are important and can increase the relic 
density by a factor of 5, in regions of parameter space where 
LSP is a high purity Higgsino state \cite{Drees2}.}. 

For large values of the {\small LSP} mass many channels are 
open, but for small values ($m_{\lsp}<40 \GeV$) only channels with  
fermions, with the exclusion of top quark, in the final state are
contributing. In these
processes the exchanged particles can be either 
a $Z$-boson and a Higgs in the $s$-channel, 
as well as a sfermion $\tilde f$ in the $t$-channel. Higgs exchanges are 
suppressed by their small couplings to light fermions, 
and sfermion exchanges are suppressed when their masses are large. Then 
the only term surviving, even for large values of squark or slepton masses, 
is the $Z$-boson exchange. 
However in the parameter region where the {\small LSP} is a 
bino, this is not coupled to the $Z$-boson resulting to very small cross 
sections enhancing dramatically the 
{\small LSP} relic density. Therefore in  
considerations in 
which the {\small LSP} is a light bino\footnote{ This happens
 when $|\mu| \gg M_W$, with $M_1 $ small 
$\approx M_W$.}, large squark 
or slepton masses are inevitably excluded, since they lead to 
large relic densities.  
If one relaxes this assumption and considers regions of the parameter space 
in which the {\small LSP} is light but is not purely bino,  
heavy squarks or sleptons may be allowed. We shall return to this later. 
Therefore the possibility for heavy $\tilde q$ or $\tilde l$ in the sparticle 
spectrum still exists, at the expense of having a light {\small LSP}
and one of the chargino states.

\begin{figure}[t] 
\begin{center} 
\epsfig{file=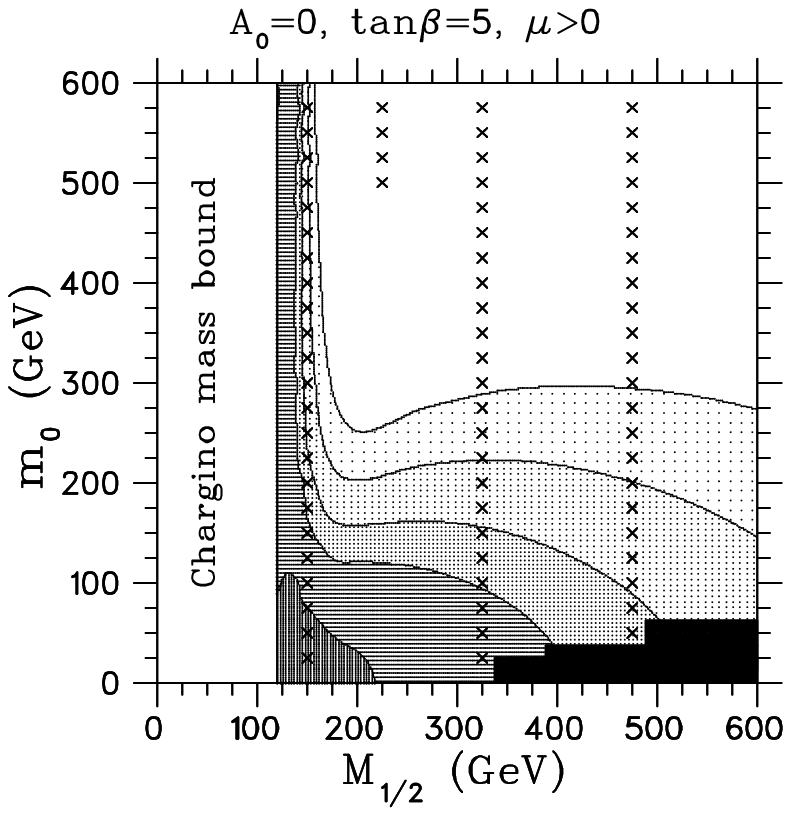,height=7.5cm,width=7.5cm} 
\hspace{.3cm}
\epsfig{file=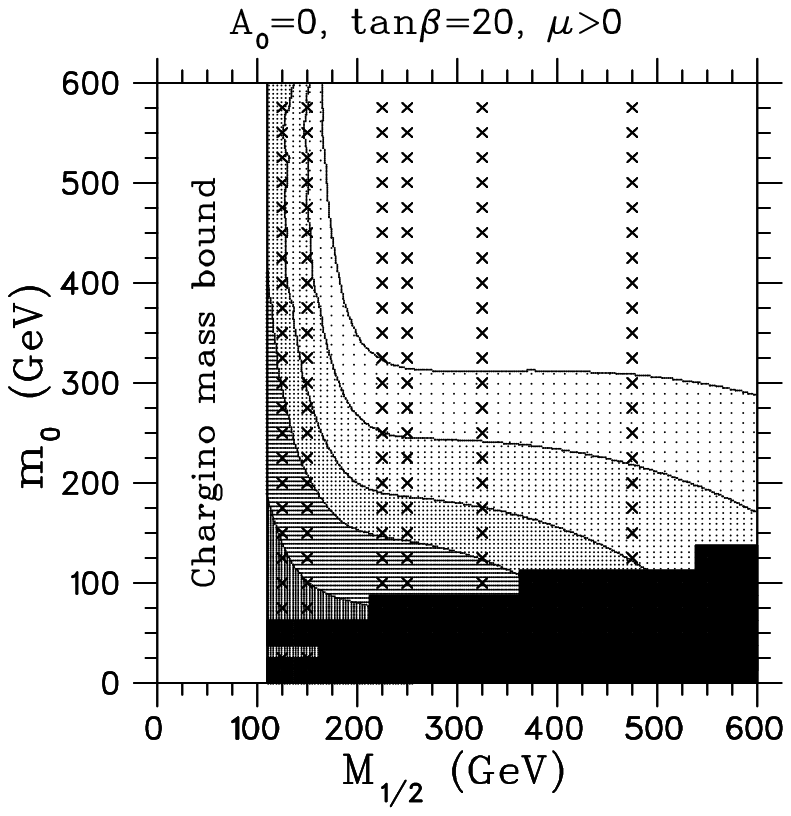,height=7.5cm,width=7.5cm}
\begin{minipage}[t]{14.cm}  
\caption[]{
The {\small LSP} relic density
$\relic$ in the ($m_0$,$M_{1/2}$) plane for
given values of $A_0$, $\tan \beta$ and sign of $\mu$. 
Grey tone regions, from darker to lighter,
designate areas in which the {\small LSP} relic density
takes values in the intervals:
$0.00 -0.08$, $0.08-0.22$,
$0.22-0.35$, $0.35-0.60$ and $0.60-1.00$ respectively.
In the blanc area $\relic > 1.0$. The dark area
corresponds to points for which the
{\small LSP} is not a neutralino. The area which
is excluded by chargino searches, 
$m_{\tilde {C}}>66\GeV$,
is also shown. Crosses denote points for which thresholds
or poles are encountered.}

\label{fig1}  
\end{minipage}  
\end{center}  
\end{figure}

We have scanned the parameter space for values of 
$m_0$, $M_{1/2}$, $A_0$ up to $1 \TeV$ and  $ \tan \beta$ 
from around 1.8 to 35 for both positive and negative values of $\mu$. 
The top quark mass is taken $175 \GeV$. 
In Figure~\ref{fig1},
we display a
representative output in the ($M_{1/2}$,$m_0$)
plane for fixed values of $A_0$ and $\tan \beta$.
Although in the displayed figure only values for $\mu>0$ are presented, in our
analysis both signs of $\mu$ have been considered.
 In the displayed figures $A_0$ takes the values 
$A_0=0$ and $\tan \beta = 5$, $20$. The five different grey tone 
regions met as we move from bottom left to right up, correspond to regions in 
which $\relic$ takes values in the intervals 
$0.00 -0.08$, 
$0.08-0.22$, $0.22-0.35$, $0.35 - 0.60$ 
and $0.60 - 1.00$ respectively\footnote{These regions have been
chosen in accord with the new bounds on $\relic$ quoted in the introduction,
and less stringent bounds on the same quantity which have been previously
used in other works.}.
 
In the blanc area covering the right up region, the relic density 
is found to be larger than unity. In the area to the left of the figure,  
the chargino mass bound is violated. Whenever a 
cross appears it designates that we are near either a pole or a threshold 
according to the criteria given previously. In these cases 
the approximations used  
are untrustworthy and no safe conclusions can be drawn. For low values 
of $M_{1/2}$ crosses correspond to mainly poles, which are either $Z$-boson 
or light Higgs, while for higher values, where 
{\small LSP} is heavier and hence new 
channels are open, these correspond to thresholds. 
For $M_{1/2} \approx 110 \GeV$ 
($\approx 90\GeV $ for $\mu < 0$) 
the lightest of the chargini has a mass close to 
its experimental lower limit. This bound is violated for all points  
$M_{1/2} \leq 100 \GeV$. The dark area at the bottom part of the figure, 
which occurs for low values of $m_0$, $m_0 \leq 100\GeV$, is excluded since
it mainly includes points for which the {\small LSP} is not a neutralino. 
In a lesser extend some of these correspond to points, which are theoretically 
excluded in the sense that  
either radiative breaking of the {\small EW} symmetry does not occur 
and/or other unwanted minima, breaking color or lepton number, are developed. 
 
For fixed $M_{1/2} > 150 \GeV$ the relic density $\relic$ 
increases as $m_0$ 
increases, due to the fact that cross sections involving sfermion 
exchanges decrease. Thus the area filled by the points corresponding to 
$\relic < 0.22$ concentrates to the left bottom of the figure 
($m_0 < 200 \GeV$). 
Also for fixed $m_0 < 250 \GeV$ the number of 
points for which $\relic < 0.22 $ decreases 
as $M_{1/2} $ increases since an increase in 
$M_{1/2} $, enlarges squark and slepton masses as well 
yielding smaller cross sections. 
If $M_{1/2} $ is further increased, for fixed $m_0$,  
the {\small LSP} will eventually 
cease to be a neutralino\footnote{The region
for which the LSP is not a neutralino depends rather strongly on the 
parameters $A_0$ and $\tan \beta$. Actually for large values of these 
parameters the light stau becomes lighter than the neutralinos.}.  
This picture changes 
for small values $M_{1/2} \approx 100 \GeV$, (see for instance the first of
Figures~\ref{fig1})
in which case small values of $ \relic < 0.22 $ can occur even 
for large values of the parameter 
$m_0 > 300 \GeV$. This happens because the 
{\small LSP} is not a bino in this region 
and $Z$-boson exchanges contribute substantially to the annihilation of 
{\small LSP}'s 
into light fermions, as explained previously. Actually in this 
region $|\mu| $ is not much larger than the {\small EW} scale $M_W$ and the 
{\small LSP}, although mostly bino, contains a sizeable portion of $H_1$
Higgsino, leading to sizeable cross sections. 
 
Therefore the neutralino relic density does not exclude 
heavy squark or slepton states, as long as {\small LSP} is light, containing  
substantial mixture 
of Higgsinos\footnote{ These corridors of low $M_{1/2}$ and large $m_0$ values 
have been also presented in Ref.~\cite{recent}. The figures of that 
reference displaying values of relic density in the ($M_{1/2}$, $m_0$) plane 
are in remarkable agreement with our corresponding figures.}.
In all cases
where this takes place, the {\small LSP} turns out to be light
$m_{\lsp} \approx 40 \GeV$ or smaller and 
thus the only open channels are those involving light fermions in the final 
state. Then the annihilation of {\small LSP}'s into neutrinos for instance, 
a channel which is always open, proceeds via the 
exchange of a $Z$-boson which is non-vanishing and dominates the reaction 
when $m_0$ is sufficiently large, due to the heaviness of sfermions. This  
puts a lower bound on $\xsec$ and hence an upper 
bound on $\relic$, which can be lower than 0.22 consistent with 
the upper experimental limits quoted in the introduction. 
On these grounds one would expect that by increasing 
$m_0$, while keeping $M_{1/2} \approx 100 \GeV$ fixed, the relic density stays 
below its upper experimental limit. However, with the exception of a few
cases, this is not so in all cases studied. In some of the cases, 
beyond a certain point along the $m_0$ axis, 
$ \relic$ exceeds the value 0.22 and starts increasing monotonically. 
This is  due to the fact that the parameter $|\mu|$ gets large again 
making {\small LSP} moving towards the bino region. In this case cross sections 
are suppressed, since the coupling of {\small LSP}
to $Z$-boson are negligible and sfermions are
quite massive, and therefore the relic density is enhanced. As a further
remark, we have to point out that the cosmologically allowed
corridors of low $M_{1/2}$ and high $m_0$ values, which we have just discussed,
are almost ruled out in view of newer data on chargino masses
{\cite{newdata}}, which push up the lower bound for the soft gaugino mass
$M_{1/2}\;$to about $130 \GeV$. 
 
It is seen from Figures~\ref{fig1}
that as $\tan \beta$ increases the points for which the 
{\small LSP} is not a neutralino increase in number. As said before,
this is due to the fact that by
increasing $\tan \beta$ the stau sfermion $\tilde{\tau}_{R}$ becomes
lighter, since masses of the sfermions of the third generation have a rather
strong dependence on $\tan \beta$. For exactly the same reasons this is also
the case when we increase the soft parameter $A_0$. For lack of space we
do not display this case \cite{LNS}.

\begin{figure}[t]  
\begin{center}  
\epsfig{file=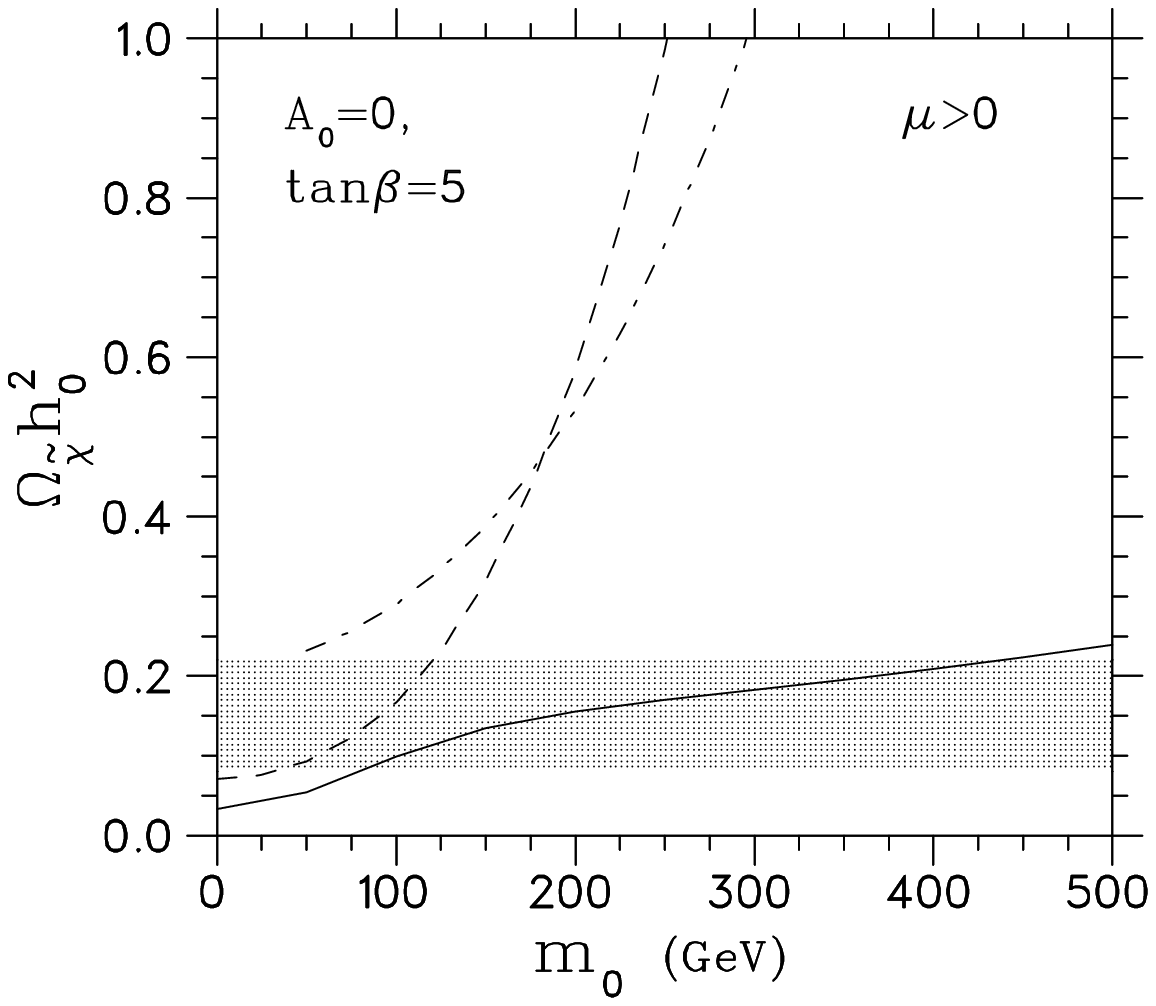,height=7.5cm,width=7.5cm} 
\hspace{.3cm}
\epsfig{file=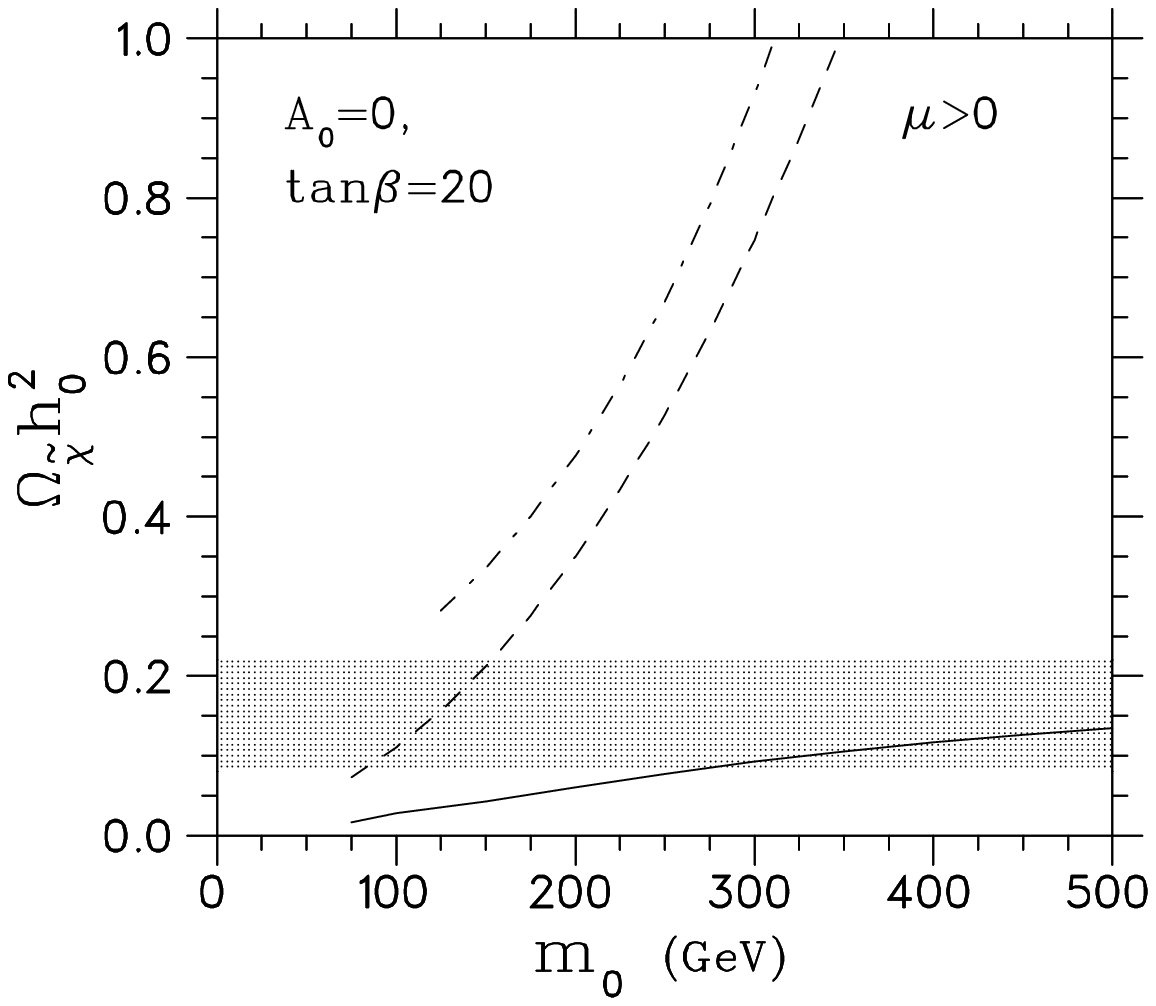,height=7.5cm,width=7.5cm} 

\begin{minipage}[t]{14.cm} 
\caption[]{The relic density as function of $m_0$ for fixed values of
$A_0$, $\tan \beta$ shown in the figure.
The solid line corresponds to $M_{1/2}=120\GeV$ ($110\GeV$)
for $\tan\beta=5$ (20), the lowest allowed
values by the chargino searches (see Figure~\ref{fig1}).
The  dashed
and dot-dashed lines correspond to $M_{1/2}=200$ and $400\GeV$
respectively. The grey area corresponds to $\relic=0.15\pm0.07$.}
\label{fig2} 
\end{minipage} 
\end{center} 
\end{figure}

\begin{figure}[b]  
\begin{center}  
\epsfig{file=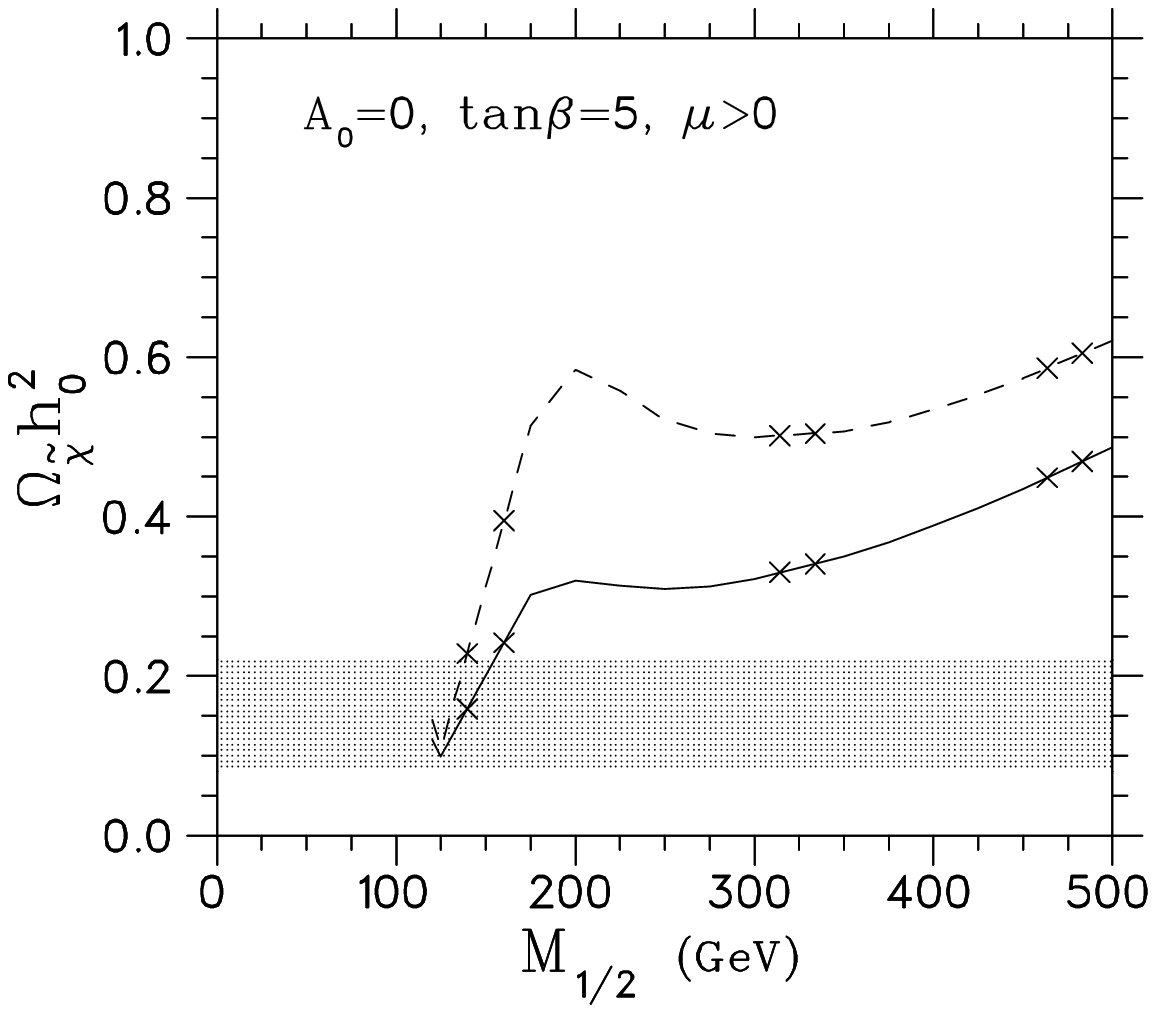,height=7.5cm,width=7.5cm} 
\hspace{.3cm}
\epsfig{file=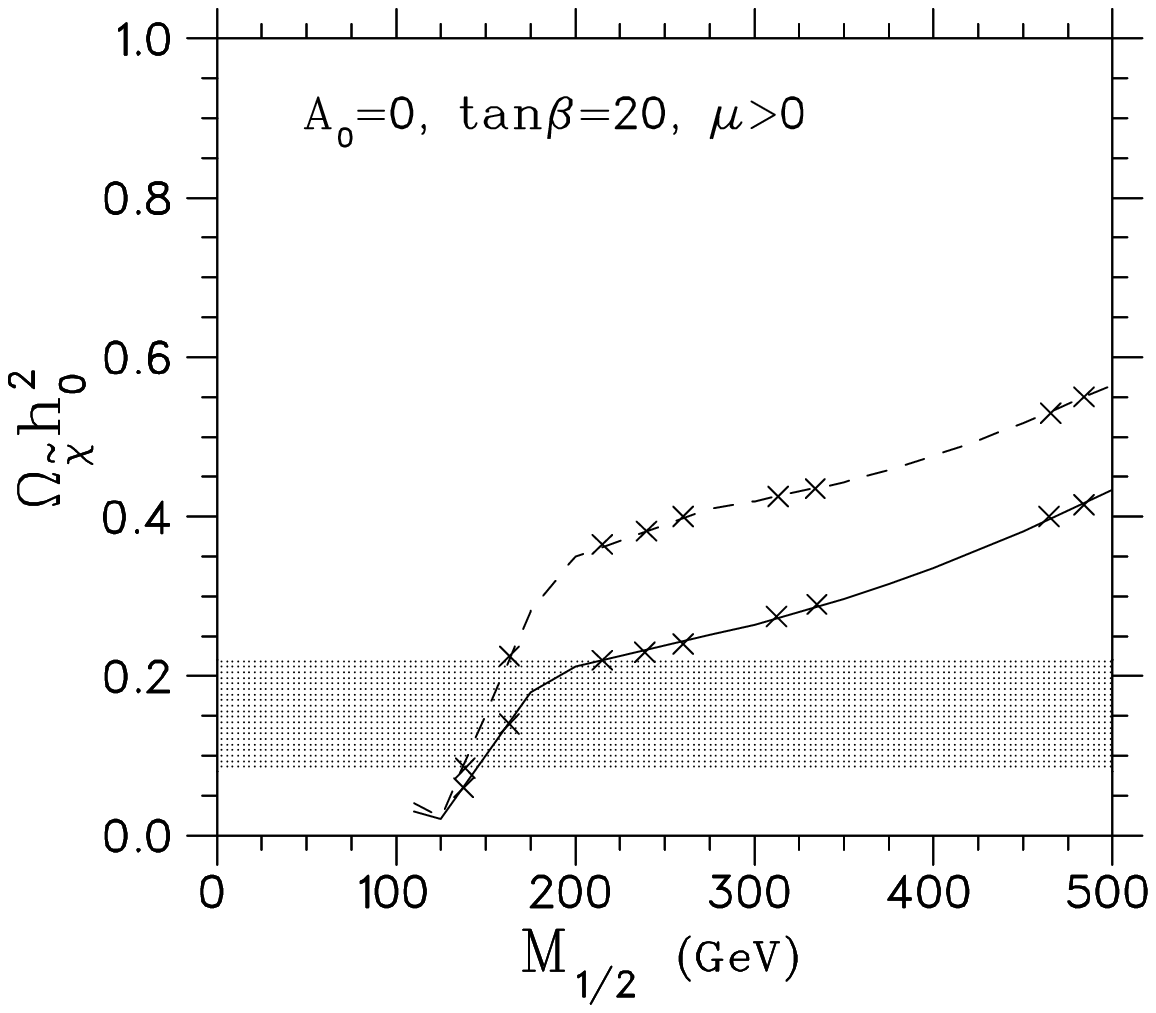,height=7.5cm,width=7.5cm} 

\begin{minipage}[t]{14.cm} 
\caption[]{The relic density as function of $M_{1/2}$ for fixed values of
$A_0$, $\tan \beta$ shown in the figure.
The solid and dashed lines 
correspond to values $m_0=150 \GeV$ and $200 \GeV$ respectively. }
\label{fig3} 
\end{minipage} 
\end{center} 
\end{figure}

\begin{figure}[t]  
\begin{center}  
\epsfig{file=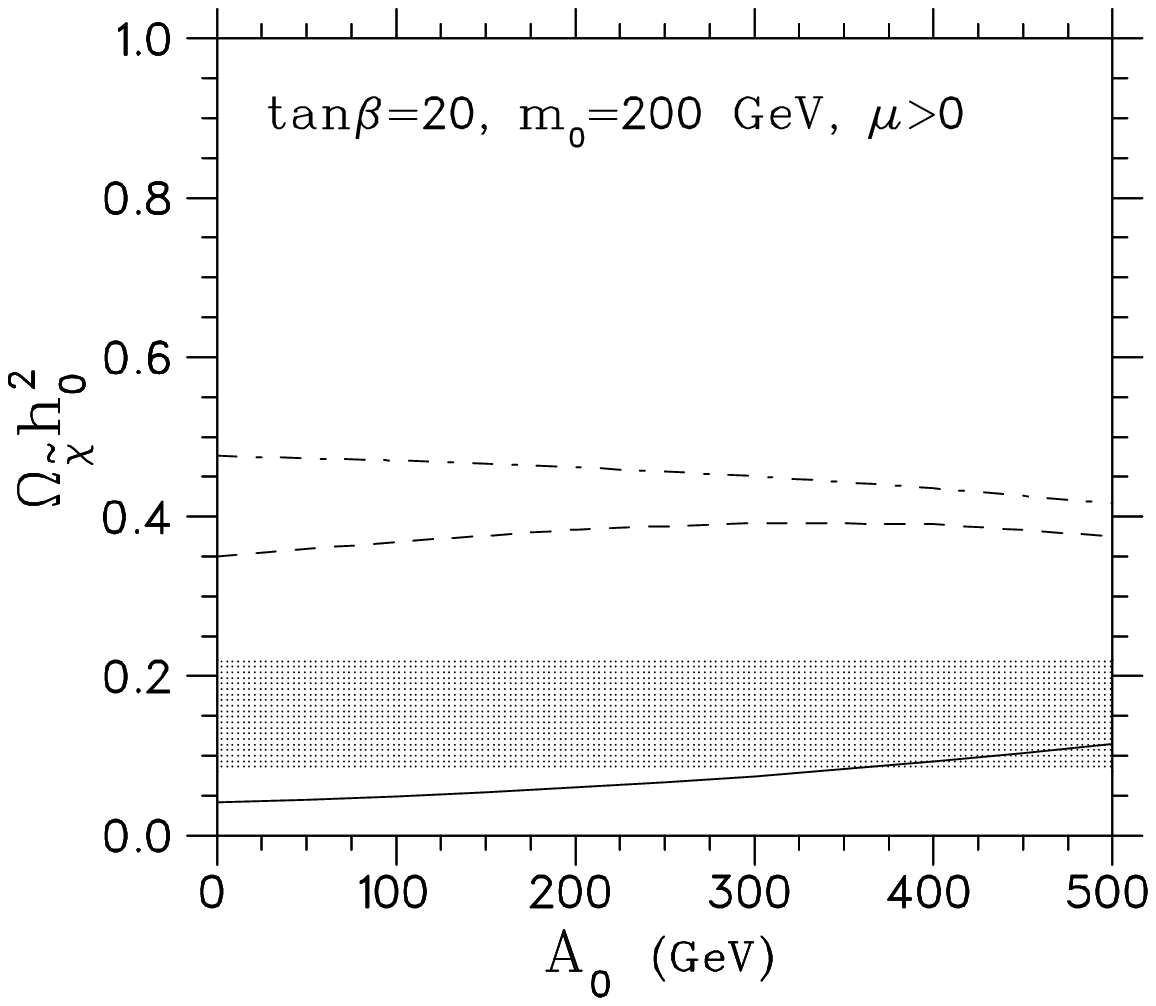,height=7.5cm,width=7.5cm} 
\hspace{.3cm}
\epsfig{file=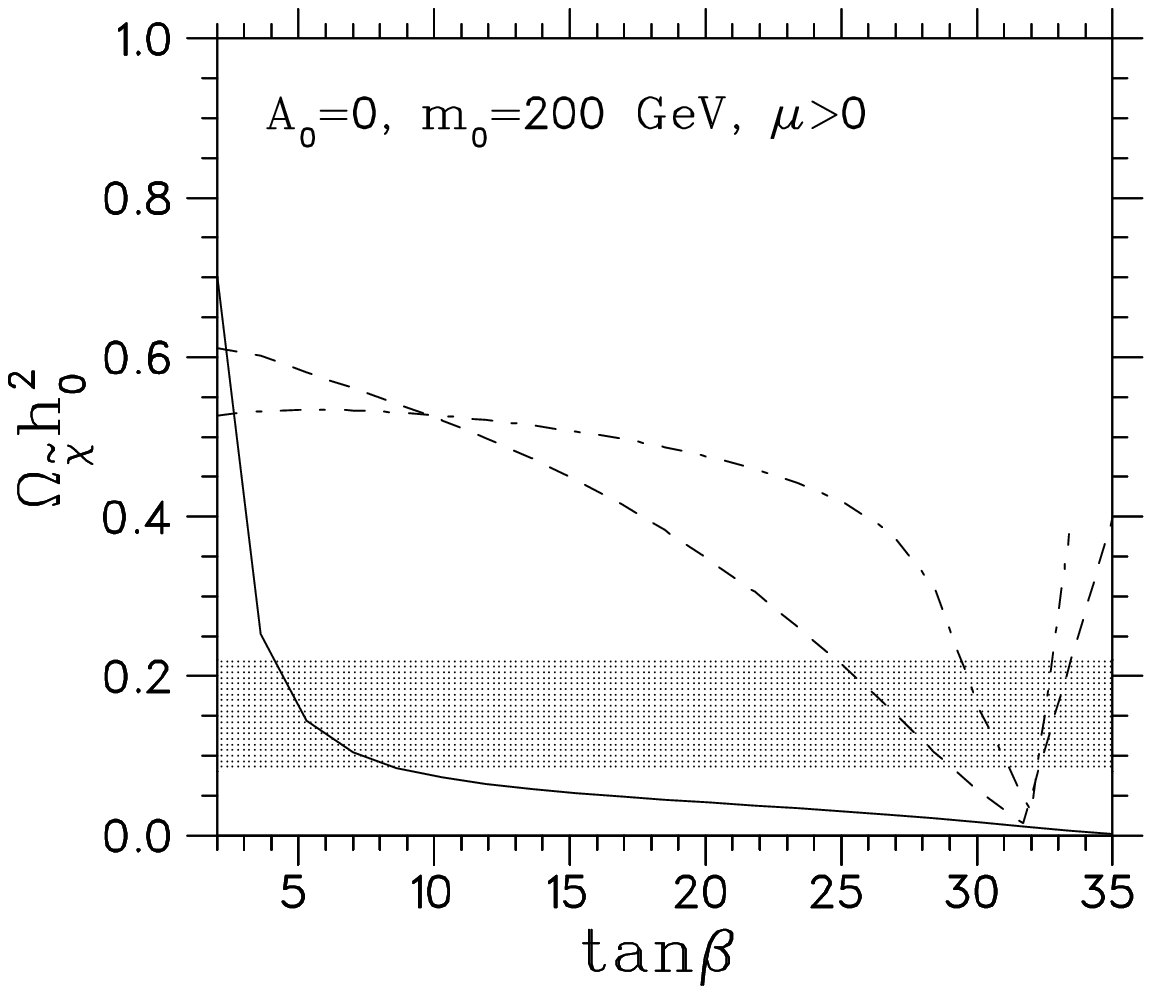,height=7.5cm,width=7.5cm} 

\begin{minipage}[t]{14.cm} 
\caption[]{The relic density as function of $A_0$ and $\tan \beta$ for
fixed values of the remaining parameters.
The  solid, dashed 
and dot-dashed lines correspond to $M_{1/2}=120$, $200$ and $400\GeV$ 
respectively.}
\label{fig4} 
\end{minipage} 
\end{center} 
\end{figure}

In Figures~\ref{fig2},\ref{fig3},\ref{fig4} we plot representative outputs
of the relic density as
function of one of the parameters $m_0$, $M_{1/2}$, $A_0$ and $\tan \beta$
keeping the other three fixed. In Figure~\ref{fig2} the relic density is
plotted against $m_0$; the lines shown correspond to different values for
$M_{1/2}$.  
Notice that at the point where each line starts 
the mass of the {\small LSP} is equal to that of ${\tilde {\tau}}_R$.
From this figure it is obvious the tendency to get
acceptable values for the relic density, as long as $m_0$ and $M_{1/2}$
are kept light. Also obvious is the fact that larger values of $m_0$ are
obtained for large values of $\tan \beta$.

In Figure~\ref{fig3}
the relic density is shown as function of the soft parameter $M_{1/2}$.
Low values of $M_{1/2}<120\GeV\;(110\GeV$) for $\tan\beta=5\;(20)$
are not allowed because of the chargino mass
bound.  It is clearly seen that low values of $M_{1/2}$ are preferred. The
upper bounds on $M_{1/2}$, which are compatible with the cosmological data,
are higher for low $m_0$ and large $\tan \beta$. In this figure crosses
denote location of poles or thresholds.

In the first of Figures~\ref{fig4} we display the relic density as a function
of $A_0$. We have chosen $\tan \beta =20$. Lowering the  value of
$\tan \beta$, keeping $m_0, M_{1/2}$ fixed,
the curves shown move upwards away from the shaded stripe, which
is cosmologically accepted. In the second figure $\relic$ is plotted as
a function of $\tan \beta $. One immediately notices the
tendency for the relic density to decrease as $\tan \beta $ gets large.

\begin{figure}[t] 
\begin{center} 
\epsfig{file=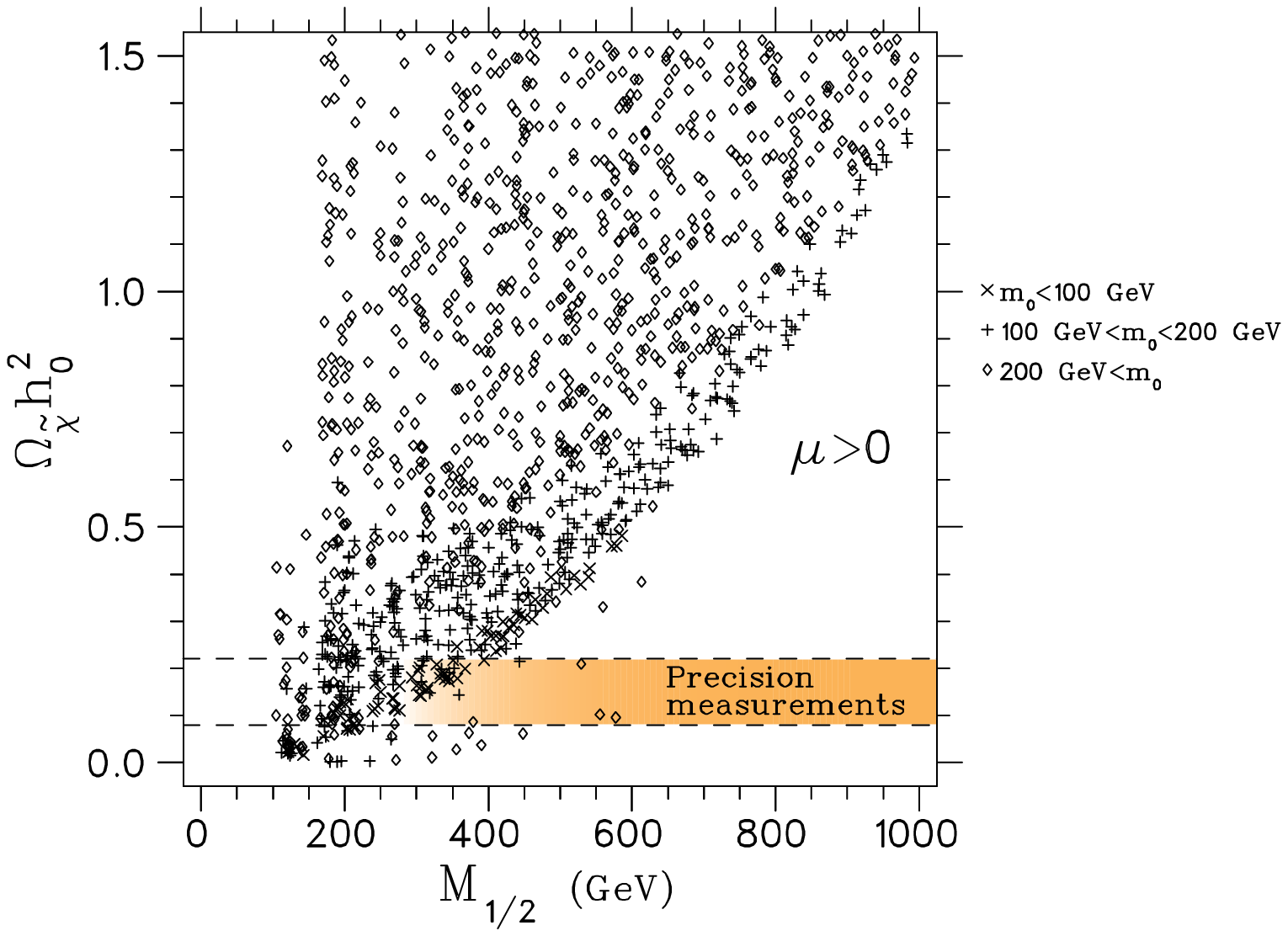,height=8.5cm,width=12.5cm} 

\vspace{.6cm}
\epsfig{file=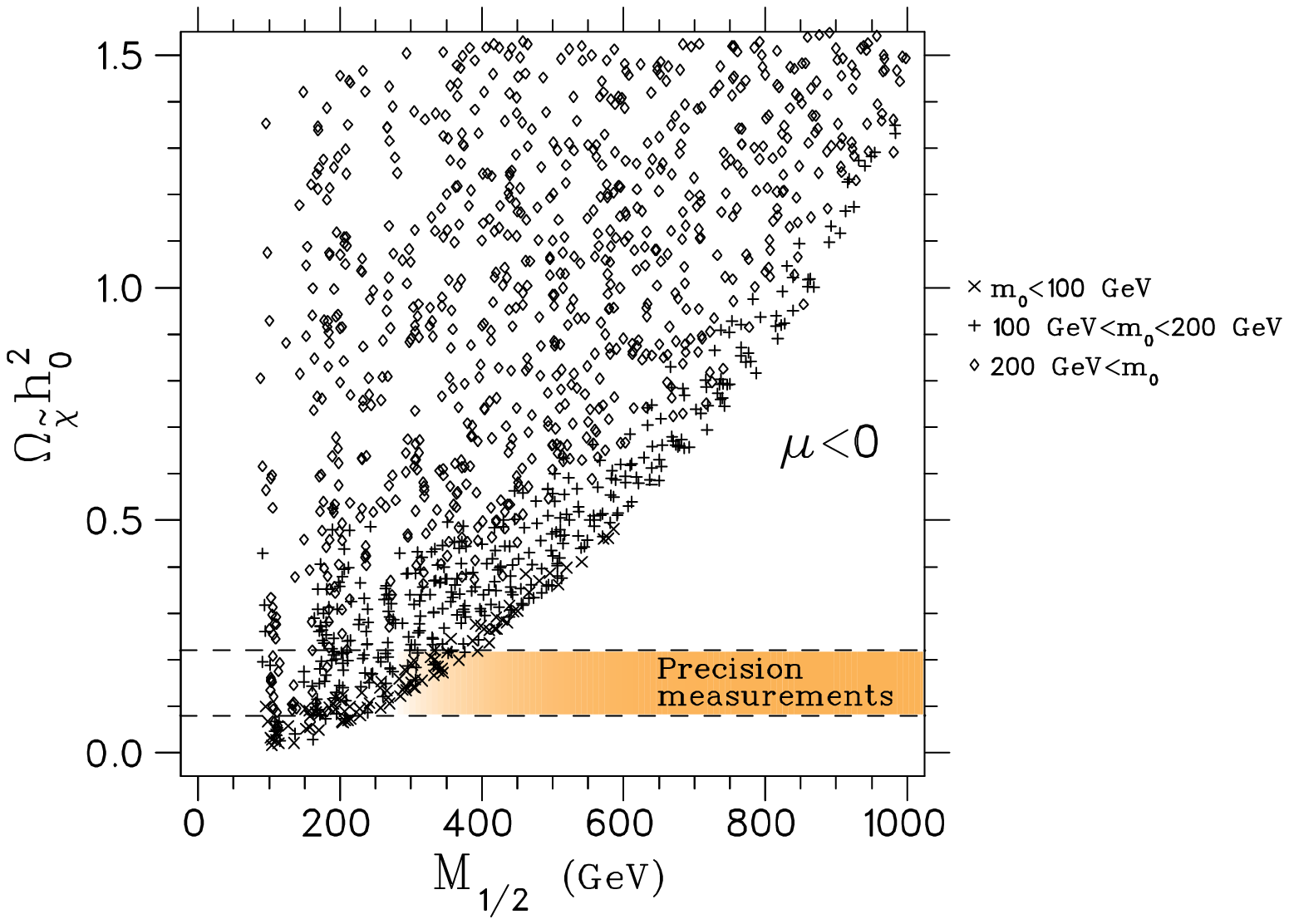,height=8.5cm,width=12.5cm} 

\begin{minipage}[t]{14.cm} 
\caption[]{Scattered plot of the relic density versus $M_{1/2}$ from a
sample of 4000 random points in the parameter space. Only the points with
$\relic$ less than 1.5 are shown. The grey tone region within the
cosmologically allowed stripe designates the region which agrees with
{\small EW} precision data. }
\label{fig5}
\end{minipage} 
\end{center}
\end{figure}

Scattered plots of {\small LSP} relic density are shown in Figure~\ref{fig5}. 
The sample consists of 4000 random points that cover the most interesting
part of the parameter space, which is within the limits :
$1.8<\tan\beta<40$, $M_{1/2}<1\TeV$, $|A_0|<500 \GeV$ and
$m_0<500 \GeV$\footnote { Higher values for $m_0$ are of relevance only
for $M_{1/2} \approx 100 \GeV$. Such low values for $M_{1/2}$
are almost ruled out by recent data. Also since $\relic$ does not depend
strongly on $A_0$ for $M_{1/2}>100\GeV$,
as it can be realised from the first of the Figures~\ref{fig4},
it suffices to focus on values $|A_0|<500 \GeV$.}.
From the given 
sample only those points which lead to relics less than 1.5 are shown in the 
figure. The experimental bounds discussed before, restrict by about 
$40 \%$ the values of the allowed points. Low values of $M_{1/2}$ 
less than about $100\GeV$ are experimentally forbidden by chargino 
searches. The points shown are striken by a cross ($\times$) when 
$m_0 < 100 \GeV$, by a plus (+) when  $100\GeV < m_0 < 200 \GeV$ 
and by a diamond ($\diamond$) when $m_0$ exceeds $200\GeV$. It is obvious 
the tendency to have $M_{1/2} < 400\GeV$ in the cosmologically interesting 
domain which lies in the stripe between the two lines at 0.08 and 0.22.
Actually 
except for a few isolated cases, all allowed points are accumulated between 
$ M_{1/2} \approx 110 - 380 \GeV$. Note that the values of the parameter 
$m_0$ in the allowed area are restricted to mainly $m_0 < 200\GeV$ 
(crosses or pluses). Only a few points with large values 
$m_0>200\GeV$ are consistent with the recent cosmological data. 
 
{\small EW} precision data are in perfect agreement with the {\small SM}
and therefore with supersymmetric extensions of it which are
characterised by a large $M_{\mathrm{SUSY}}$.
In fact overall fits to the {\small EW} precision data show
a preference towards large values of $M_{\mathrm{SUSY}}$,
in which case better fits (lower $\chi^2$) are obtained \cite{Lahanas}.
In the constrained {\small MSSM} lower bounds on
$M_{1/2}$ can be established from the experimental value of the
effective weak mixing angle $\sin^2\theta_{\mathrm{eff}}$,
measured in {\small SLC} and {\small LEP} experiments.
If the combined {\small SLC} and {\small LEP} data are used,
$M_{1/2}$ is restricted to be larger than about $300\GeV$ (see
Dedes {\em et. al.} in Ref.~\cite{Lahanas}\footnote{SLC data
alone leave more freedom by allowing for lower
$M_{1/2}$ values. On the contrary  LEP data
alone prefer large values of $M_{1/2}$.}).
This lower bound on $M_{1/2}$ can be further increased by about
$200\GeV$, if in addition gauge coupling unification 
is assumed\footnote{However since this result depends 
sensitively on possible existence of High Energy thresholds we do not
impose such a strong lower bound on $M_{1/2}$ values.}.

Following the previous discussion, in Figure~\ref{fig5} we have 
shaded in grey the region which is consistent with 
{\small EW} precision data. This starts at about $300\GeV$
and progressively becomes darker as we move to larger
$M_{1/2}$ values, where the {\small SM} limit is attained and
better agreement with experimental data is obtained.
 We observe that
only a few points, the majority of them being very close to $400\GeV$, are
compatible with both astrophysical and {\small EW} precision measurements.
The values of $M_{1/2}$ and $m_0$ for which theoretical predictions agree
with both data, are constrained within the regions
$300 \GeV<M_{1/2}<400\GeV$ and $m_0<200\GeV$. Performing a more
refined scanning in the parameter space with values of $M_{1/2}$, $m_0$ in
the aforementioned range and varying the other two in the range
$1.8<\tan\beta<40$, $|A_0|<1\TeV$, we have found the following lower and
upper bounds on the masses of the {\small LSP}, the lighter of charginos, staus, 
stops and the light scalar Higgs: 
\bea  
m_{\mathrm{LSP}}\;&:&\; 113\GeV\;-\;149\GeV\,,\nonumber \\ 
m_{\tilde{C}}\;\;\;\;&:&\; 209\GeV\;-\;278\GeV\,,\nonumber \\ 
m_{{\tilde{\tau}}_R}\;\;\;&:&\;118\GeV\;-\;168\GeV,\nonumber \\ 
m_{{\tilde{t}}_1}\;\;\;&:&\;390\GeV\;-\;720\GeV\,,\nonumber \\ 
m_{{h}_0}\;\;\;&:&\;76\GeV\;-\;122\GeV\,.\nonumber  
\eea  
These refer to the case $\mu>0$. Analogous bounds are found for 
$\mu<0$. With the exception of the Higgs lower bound, which is 
increased by about $20\GeV$, the bounds for the remaining sparticles 
in the $\mu<0$ case are almost unchanged. Since we have neglected
coannihilation effects, the conclusions reached are actually valid outside
the stripe
$m_{\lsp}<m_{{\tilde{\tau}}_R} \lesssim 1.2 \; m_{\lsp} $. Inside this
band $\lsp - {\tilde{\tau}}$ coannihilations dominate the cross sections,
decreasing $\lsp$ relic densities leaving corridors of opportunity to
high $M_{1/2}$ and $m_0$ values as emphasized in other studies \cite{Falk1}.

In conclusion we can say that the new stringent bound on the 
matter relic density $\relic \lesssim 0.22$ extracted from recent data,
prefer low supersymmetry breaking scales and hence a light sparticle
spectrum, provided $m_{{\tilde{\tau}}_R} > 1.2 \; m_{\lsp}$. 
This behaviour is exactly the opposite of what happens in the
{\small EW} precision measurements physics. Reconciling 
{\small LEP}/{\small SLC} data with recent cosmological measurements
restricts the parameters of the constrained minimal supersymmetry, yielding
values of some of the sparticle masses not far away from their lower
experimental limits.

\vspace{3cm}
\noindent 
{\bf Acknowledgments} \\ 
\noindent 
A.B.L. acknowledges support from ERBFMRXCT--960090 TMR programme
and D.V.N. by D.O.E. grant DE-FG03-95-ER-40917.  
V.C.S. acknowledges an enlightening discussion with D.~Schwarz.

\newpage
%%%%%%%%%%%%%%%%%%%%%%%%%% Figure 1 %%%%%%%%%%%%%%%%%%%%%%%%%%%%%%%% 

\newpage
%%%%%%%%%%%%%%%%%%%%%%%%%% Figure 2 %%%%%%%%%%%%%%%%%%%%%%%%%%%%%%%%  
%%%%%%%%%%%%%%%%%%%%%%%%%%%%%%%%%%%%%%%%%%%%

%%%%%%%%%%%%%%%%%%%%%%%%%% Figure 3 %%%%%%%%%%%%%%%%%%%%%%%%%%%%%%%%  
%%%%%%%%%%%%%%%%%%%%%%%%%%%%%%%%%%%%%%%%%%%%

\newpage 
%%%%%%%%%%%%%%%%%%%%%%%%%% Figure 4 %%%%%%%%%%%%%%%%%%%%%%%%%%%%%%%%  
%%%%%%%%%%%%%%%%%%%%%%%%%%%%%%%%%%%%%%%%%%%%

\newpage
%%%%%%%%%%%%%%%%%%%%%%%%%% Figure 5 %%%%%%%%%%%%%%%%%%%%%%%%%%%%%%%% 

\end{document}